\documentclass[
aps,prc,groupedaddress,showpacs,nofootinbib,showkeys]{revtex4}
\usepackage{graphicx}
\usepackage{color}
\usepackage{amsmath}
\usepackage{amssymb}
\begin{document}
\title{Hyperons and massive neutron stars: vector repulsion and SU(3) symmetry}
\date{Preprint submitted to Phys. Rev. C on April 2. 2012}
\author{S.~Weissenborn}
\email{s.weissenborn@thphys.uni-heidelberg.de}
\author{D.~Chatterjee}
\email{d.chatterjee@thphys.uni-heidelberg.de}
\author{J.~Schaffner-Bielich}
\email{schaffner@thphys.uni-heidelberg.de}
\affiliation{Institut f{\"u}r Theoretische Physik, University of Heidelberg,\\
Philosophenweg 16, D-69120 Heidelberg, Germany}
\keywords{neutron stars, equation of state, hypernuclei, hadronic matter}
\pacs{97.60.Jd, 26.60.Kp, 21.80.+a, 21.30.Fe}
\begin{abstract}
With the discovery of massive neutron stars such as PSR J1614-2230, the question has arisen
whether exotic matter such as hyperons can exist in the neutron star core. We examine the conditions under which 
hyperons can exist in massive neutron stars. We consistently investigate the vector meson-hyperon coupling, going from
SU(6) quark model to a broader SU(3) symmetry. We propose that the maximum neutron star mass decreases linearly with the
strangeness content $f_s$ of the neutron star core as $M_{max}(f_s) = M_{max}(0) - 0.6 M_{\odot} (f_s/0.1)$,
which seems to be independent of the underlying nuclear equation of state and the vector baryon-meson coupling scheme.
Thus, pulsar mass measurements can be used to constrain the hyperon fraction in neutron stars. 
\end{abstract}
\maketitle
\section{Introduction}\label{intro}
The discovery of the massive neutron star PSR J1614-2230 has raised new challenges for 
theories of dense matter beyond nuclear saturation density. Shapiro delay measurements from radio timing observations of the binary millisecond pulsar indicate
a large mass of 1.97$\pm$0.04$M_{\odot}$ of the neutron star \cite{Demorest10}. The core of a neutron star harbors a dense matter
environment, which could be the site for strangeness containing matter, such as hyperons. 
Though nuclear interactions in the saturation regime are well understood, one has to utilize neutron star observations to find clues about the 
physics of cold and dense matter beyond several times saturation density. Any theory of ultra dense matter has to explain the recently observed large neutron star mass.
According to existing models of dense matter, the presence of hyperons leads  to a considerable softening of the equation of state (EoS), 
resulting in a corresponding reduction of the maximum mass of the neutron star \cite{Glendenning92derivativecoupling,Baldo00,Vidana00,Djapo}.
Then the existing theories involving hyperons are in conflict with the large pulsar masses \cite{Lattimer10}. 
On including hyperons, most relativistic models obtain maximum neutron star masses in the 
range $1.4-1.8 M_{\odot}$ \cite{Glen85,GM1,Knorren,BalbergGal,PalHanauske,Hanauske00,Zschiesche,Long12}. 
However, in exceptional cases, neutron stars with maximum masses larger than 2$M_{\odot}$ have been obtained, either by pushing the threshold
for appearance of hyperons to higher densities, or due to strong hyperon vector repulsion \cite{Long12,Huber,Hofmann,Rikovska,Dhiman07,Dexheimer08,Bombaci08,Cavagnioli11}.
Taurines et al. \cite{Taurines} achieved large neutron star masses including hyperons by considering a model with density dependent coupling constants,
which were varied nonlinearly with the scalar field. Recently, Bednarek et al. \cite{Bednarek2011} 
also achieved a stiffening 
of the EoS by using a non-linear relativistic mean field (RMF) model with quartic terms involving the hidden strangeness vector meson. 
In addition to the inclusion of such a meson into a density dependent RMF model, 
Lastowiecki et al. \cite{Lastowiecki12} assumed a quark matter core in order to obtain massive stars. 
Bonanno and Sedrakian \cite{Bonanno12} also modeled massive neutron stars with a hyperon and quark core 
using a fairly stiff EoS and vector repulsion among quarks. 
In several studies, the maximum neutron star masses obtained when including hyperons were 
not very different from those containing nucleons only \cite{Hofmann,Rikovska,Dexheimer08}. 
In more sophisticated models such as the Brueckner-Hartree-Fock model, the maximum neutron star masses were generally found to be lower than $1.6M_{\odot}$ 
which is in contradiction with observed pulsar masses \cite{Baldo00,Vidana00,Djapo,Baldo98,Nishizaki,SchulzeVidana,SchulzeRijken,Logoteta12}.\\
From the studies cited in the previous paragraph, it seems that the possible presence of hyperons in massive neutron 
stars is in many cases reconciled by incorporating large vector repulsion in an ad hoc way. 
In contrast, we investigate the role of vector repulsion starting from symmetry arguments. 
Assuming SU(3) symmetry, we perform a controlled parameter study and constrain the parameters using the observed mass of PSR J1614-2230. 
This procedure is in line with modern microscopic models for realistic baryon-baryon potentials such as the Bonn potentials \cite{Machleidt87} and 
the Nijmegen potentials \cite{Rijken99} which adopt SU(3) symmetry to describe the baryon interactions for the baryon octet. 
For our investigations, we employ a RMF model, in which the parameters
are calibrated around nuclear saturation density \cite{Glen85,Knorren,Schaffner96}. However, the extrapolation of such properties
to supranuclear densities presents uncertainties. In a previous paper \cite{WsbChtt12}, we investigated how the uncertainty in nuclear saturation properties,
such as effective nucleon mass or nuclear compressibility, or hypernuclear properties, such as potential depths
of hyperons in nuclear matter, could influence our conclusions about the presence of hyperons in the core of
massive neutron stars. In this work, we question the fundamental assumption of SU(6) symmetry, which relates the
hyperon couplings to the nuclear couplings.\\ 
\indent This paper is 
organized in the following way. In Sec. \ref{model}, we describe the model to calculate the 
EoS. The parameters of the model are listed in Sec. \ref{parameters}. The results of our calculations are
discussed in Sec. \ref{results}, and a summary of our conclusions is given in Sec. \ref{summary}. \\
\section{Theoretical Model}\label{model}
 As elaborated in our previous paper \cite{WsbChtt12}, a RMF theoretical model 
is adopted to describe neutron star matter subject to chemical 
equilibrium and charge neutrality. The octet baryons, electrons and muons
are considered as constituents of the core. In this model \cite{Schaffner96}, baryon-baryon interaction is mediated by the exchange
of scalar ($\sigma$), vector ($\omega$) and isovector ($\rho$) mesons. The hyperon-hyperon interaction
is incorporated through additional strange scalar ($\sigma^*$) and vector ($\phi$) mesons. The stiffest possible
EoS within the model is obtained on inclusion of the strange vector meson $\phi$ and by omitting the $\sigma^*$ meson, and
is referred to as the ``model $\sigma \omega \rho \phi$`` \cite{WsbChtt12}.
The vector meson-hyperon couplings in this model are related to those of nucleons through the symmetry of the 
SU(6) quark model. In this paper a consistent investigation for the vector meson coupling is achieved, going from the
SU(6) symmetry to the more general SU(3) symmetry, in order to provide a general analysis for the role of repulsive hyperon
interactions on the properties of neutron stars.
\subsection{Determination of vector meson-hyperon couplings} \label{sechypcoup}
\subsubsection{Flavor SU(3) and the baryon-meson interaction Lagrangian}
SU(3) in flavor space can be regarded as a symmetry group of strong interaction 
(restricting it to three quark flavors, up, down and strange). 
For neutron stars, we only consider the $[8]$ representation for the baryons, namely the $J^P=\frac{1}{2}^+$ octet. 
 The product of the baryon and meson representations reads:
\begin{equation}
 [8]\otimes[8]=[1]\oplus[8]_A\oplus[8]_S\oplus[10]\oplus[27].
\end{equation}
With the help of matrix representations for the baryon octet ($B$) and meson nonet (singlet state $M_1$ and octet states $M_8$), SU(3) invariant expressions 
can be constructed \cite{Rijken99}. The interaction Lagrangian for the whole meson nonet and the baryons can be written as a sum of terms, 
one coming from the coupling of the meson singlet to the baryon octet (S), 
and other two terms from the interaction of the meson octet and the baryons - one being antisymmetric (F)
 and the other being symmetric (D) \cite{deSwart63,deSwart65}:
\begin{equation}
   {\cal L}_{\it{int}} = - g_{8}\sqrt{2}
     \left[\alpha Tr \left( [\overline{B},M_8] B \right) +
     (1-\alpha) Tr \left( \{\overline{B},M_8 \} B \right) \right]  
     - g_{1}{\sqrt{\frac{1}{3}}} 
     Tr (\overline{B}B) Tr(M_1)\:\:\:.          \label{LIsu3}
\end{equation}

Here, $g_8$ and $g_1$ denote the meson octet and singlet coupling constants. The  $F/(F+D)$ ratio $\alpha$, which lies in the range 
of $0\leq\alpha\leq1$, is a weight factor for the contributions of the symmetric $D$ and the antisymmetric $F$ couplings relative to 
each other. \\
The assumption of SU(3) symmetry implies that all possible combinations of couplings for each type of meson
with all possible baryons can be described
with only four parameters: the singlet coupling constant $g_1$, 
the octet coupling constant $g_8$, the $F/(F+D)$ ratio $\alpha$ \cite{Rijken99,Schaffner94} 
and a "mixing angle" $\theta$ relating the physical isoscalar mesons to their pure octet and singlet counterparts.
The $SU(6)$ quark model is obtained as the combination of flavor $SU(3)$ with spin $SU(2)$ 
(and is thus a special case of flavor $SU(3)$).

We restrict our attention to vector coupling terms containing only baryons of the same 
species \cite{Dover85} and denote the $F/(F+D)$ ratio corresponding to vector meson interaction by $\alpha_V$. 
The requirement of spin independence for the couplings of the identically flavored 
$\Sigma$ and $\Lambda$ baryons:
\begin{equation}
 g_{\it{\Sigma\omega}}=g_{\it{\Lambda\omega}}, \> g_{\it{\Sigma\phi}}=g_{\it{\Lambda\phi}}
\end{equation}
leads to a fixed value of the $F/(F+D)$ ratio $\alpha_V$:
\begin{equation}
 \alpha_V=1\:\:\:.
\end{equation}
This means, we have a pure $F$-type coupling in the interaction Lagrangian (\ref{LIsu3}).
Further, the proposition that the nucleon does not couple to the $\phi$-meson which  
is a pure $s \bar{s}$-state (i.e. $g_{N \phi}=0$), in the case of "ideal mixing" for the $\omega$ and $\phi$ meson 
\cite{Nakamura},
\begin{equation}
 {\it{tan}}\theta_V=\frac{1}{\sqrt{2}}\label{idealmixing}\:\:\:,
\end{equation}
leads to the relative coupling strengths \cite{Schaffner94}
\begin{equation}
 g_{N\omega}:g_{\Lambda\omega}:g_{\Sigma\omega}:g_{\Xi\omega}
=3:2:2:1
\label{omegacouplingssu6}
\end{equation} 
and
\begin{equation}
g_{\Lambda\phi}:g_{\Sigma\phi}:g_{\Xi\phi}=1:1:2
\label{phicouplingssu6}
\end{equation}
where equations (\ref{omegacouplingssu6}) and (\ref{phicouplingssu6}) are related through \cite{Dover85}:
\begin{equation}
 g_{\it{\Lambda\omega}}= - \frac{1}{\sqrt{2}}g_{\it{\Xi\phi}}\label{omegaphisu6}\:\:\:.
\end{equation}
This leaves only one degree of freedom, say $g_{\it{N\omega}}$ which is fitted to 
saturation properties of nuclear matter \cite{GM1}. 
We fit the $\sigma$ meson hyperon couplings to the potential depths of hyperons in nuclear matter (see \cite{WsbChtt12} for
discussions). 
\subsubsection{Beyond SU(6)}\label{beyondsu6}
In order to explain the large observed mass of PSR J1614-2230 we might either introduce 
more particles and parameters into our RMF model, or reconsider the necessity of fixing them to the $SU(6)$ values.\\
\indent From the quadratic mass formula for the mesons one obtains $\theta_V\approx40^o$ which 
is quite close to the ideal mixing angle $\theta\approx35.3^o$ \cite{Nakamura}. 
Thus, we may retain the condition of ideal mixing for the vector mesons. 
The parameters $\alpha_V$, $g_1$ and $g_8$ however, we may consider as being free. 
Since we later fix $g_{\it{N\omega}}$ to saturation properties of nuclear matter, we combine the singlet 
and octet coupling constants to a single parameter $z$ which we define as
\begin{equation}
 z:= \frac{g_8}{g_1}\label{definez}
\end{equation}
and keep the parameters $\alpha_V$, $z$ and $g_{\it{N\omega}}$. 
If we set $z$ to its SU(6) value $z=1/\sqrt{6}$ and use ideal mixing while varying $\alpha_V$, 
which by definition lies within the range  $0\leq\alpha_V\leq1$, it gives:
\begin{eqnarray}
\begin{array}{ccccccc}
  \displaystyle\frac{g_{\it{\Lambda\omega}}}{g_{\it{N\omega}}}=\frac{2\alpha_V+4}{4\alpha_V+5}&&&
 \displaystyle\frac{g_{\it{\Sigma\omega}}}{g_{\it{N\omega}}}=\frac{8-2\alpha_V}{4\alpha_V+5}&&&
 \displaystyle\frac{g_{\it{\Xi\omega}}}{g_{\it{N\omega}}}=\frac{5-2\alpha_V}{4\alpha_V+5}
\end{array}\nonumber
\end{eqnarray}

\begin{equation}
\displaystyle\frac{g_{\it{N\phi}}}{g_{\it{N\omega}}}=\sqrt{2}\cdot\frac{4\alpha_V-4}{4\alpha_V+5}  \label{alphafreecouplingsnew}
\end{equation}

\begin{eqnarray}
\begin{array}{ccccccc}
\displaystyle\frac{g_{\it{\Lambda\phi}}}{g_{\it{N\omega}}}=\sqrt{2}\cdot\frac{2\alpha_V-5}{4\alpha_V+5} &&&
\displaystyle\frac{g_{\it{\Sigma\phi}}}{g_{\it{N\omega}}}=-\sqrt{2}\cdot\frac{2\alpha_V+1}{4\alpha_V+5} &&&
\displaystyle\frac{g_{\it{\Xi\phi}}}{g_{\it{N\omega}}}=-\sqrt{2}\cdot\frac{2\alpha_V+4}{4\alpha_V+5}~.
 \end{array}\nonumber
\end{eqnarray}
If we set $\alpha_V$ to its SU(6) value $\alpha_V=1$ and use ideal mixing while keeping $z$ as a free parameter instead we arrive at:
\begin{eqnarray}
\begin{array}{ccccccc}
\displaystyle\frac{g_{\it{\Lambda\omega}}}{g_{\it{N\omega}}}=\frac{\sqrt{2}}{\sqrt{2}+\sqrt{3}z}=\frac{g_{\it{\Sigma\omega}}}{g_{\it{N\omega}}} &&&&&&
\displaystyle\frac{g_{\it{\Xi\omega}}}{g_{\it{N\omega}}}=\frac{\sqrt{2}-\sqrt{3}z}{\sqrt{2}+\sqrt{3}z}\\
\end{array}\nonumber
\end{eqnarray}
\begin{eqnarray}
\begin{array}{ccccccc}
\displaystyle\frac{g_{\it{N\phi}}}{g_{\it{N\omega}}}=\frac{\sqrt{6}z-1}{\sqrt{2}+\sqrt{3}z} &&&
\displaystyle\frac{g_{\it{\Lambda\phi}}}{g_{\it{N\omega}}}=\frac{-1}{\sqrt{2}+\sqrt{3}z}=\frac{g_{\it{\Sigma\phi}}}{g_{\it{N\omega}}} &&&
\displaystyle\frac{g_{\it{\Xi\phi}}}{g_{\it{N\omega}}}=-\frac{1+\sqrt{6}z}{\sqrt{2}+\sqrt{3}z}~.
 \end{array}\label{zfreecouplings}
\end{eqnarray}
If we require the interaction due to $\omega$ exchange to be repulsive for all baryons, we want no changes of sign in the $\omega$ couplings, 
i.e. especially  $g_{\it{\Xi\omega}} / g_{\it{N\omega}} \geq 0$ and therefore $0\leq z\leq 2/\sqrt{6}$. 
The couplings of the $\rho$ meson to the baryons may be related to $g_{N\omega}$ by 
relations analogous to (\ref{alphafreecouplingsnew}), (\ref{zfreecouplings}) involving $\alpha_V$ or z. 
However, since the $\rho$ couplings control the asymmetry energy and its density dependence L, $g_{N\rho}$ may alternatively be fitted 
directly to the asymmetry energy coefficient at saturation (following Ref. \cite{GlendenningBook}). For the hyperons, the $\rho$ coupling strengths 
are then related to $g_{N\rho}$ according to isospin. 
This method we adopt for our calculations and thereby we avoid the problem of having $g_{N\rho}=0$ at $z=0$, $\alpha_V=1$ which 
would yield unphysically low values for the symmetry energy coefficient. 
For the two ways to fix the $\rho$ couplings, the difference in the obtained 
maximum neutron star masses is rather negligible ($\Delta M_{\it{max}}\lesssim0.05 M_\odot$).\\
From the relations (\ref{alphafreecouplingsnew}) and (\ref{zfreecouplings}) it is clear that for 
values of $\alpha_V$ or z not coinciding with SU(6) the $\phi$ meson couples to the nucleon which is supported by the large strange quark condensate in the nucleon found in lattice 
gauge simulations \cite{Dong94,Ellis95}.
\section{Parameters of the model}\label{parameters}
\subsection{Nucleon-Meson coupling constants}
The nucleon-meson coupling constants $g_{\sigma N}$, $g_{\omega N}$, $g_{\rho N}$, $b$ and $c$ are determined from the saturation properties of nuclear matter \cite{GlendenningBook}:
 binding energy $ =-16.3$ MeV, baryon density $n_0=0.153$ fm$^{-3}$, and asymmetry energy coefficient $a_{\rm asy}=32.5$ MeV.
The incompressibility $K$ and effective nucleon mass $m^*_N/m_N$ are varied according to the parameter set in consideration 
(e.g. GM1 \cite{GM1}, NL3 \cite{NL3}, TM1 \cite{Sugahara94}). 
\subsection{Hyperon-Meson coupling constants}
 The strange and non-strange vector meson-hyperon couplings have already been described in detail in the previous section. 
The non-strange scalar meson-hyperon couplings are fitted to the potential depths for the hyperons in nuclear matter. The 
following values $U_\Lambda^{(N)}=-30$ MeV, $U_\Sigma^{(N)}=+30$ MeV, $U_\Xi^{(N)}=-28$ MeV have been adopted 
from the hypernuclear experimental data (see e.g. \cite{Dover82,Batty94,Mares95,Harada05}, a summary of usual choices is found in \cite{Schaffner00}). 
However, the particular choice of hyperon potentials does not have crucial consequences regarding the maximum mass 
of neutron stars as has been discussed in \cite{WsbChtt12}. In fact, a larger $\Xi$ potential 
is the more conservative choice as it gives a slightly smaller maximum mass.\\
\indent When fixing the parameter $g_{N \omega}$ we remember that by means of non-vanishing $g_{N\phi}$ the $\phi$ meson 
also contributes to the saturation properties of nuclear matter. In particular, in all thermodynamic quantities, 
e.g. in the nucleon chemical potential, we have to replace the term 
\begin{equation}
 \tilde{g}_{\it{N\omega}}\tilde{\omega} \longrightarrow g_{\it{N\omega}}\omega+g_{\it{N\phi}}\phi\label{substitutegparagnew}
\end{equation}
where the tilde ( $\tilde{\:}$ ) denotes the coupling and corresponding field at nuclear saturation for the case of $g_{N\phi}=0$. 
RMF models without vector meson self-interaction (e.g. GM1, NL3) yield as equations of motion for the $\omega$ and $\phi$:
\begin{equation}
 g_{\it{N\omega}}\omega=\frac{g_{\it{N\omega}}^2}{m_\omega^2}\rho_N\:\:\:\:\:\:,\:\:\:\:\:\:\:
g_{\it{N\phi}}\phi=\frac{g_{\it{N\phi}}^2}{m_\phi^2}\rho_N\label{omegaandphirmfeom}\:\:\:.
\end{equation}
This allows to rewrite the substitution prescription (\ref{substitutegparagnew}) as
\begin{equation}
 \tilde{g}_{\it{N\omega}}^2 \longrightarrow g_{\it{N\omega}}^2\left(1+\frac{g_{\it{N\phi}}^2}{g_{\it{N\omega}}^2}\frac{m_\omega^2}{m_\phi^2}\right) \label{gnewzfree}
\end{equation}
where the ratio $g_{\it{N\phi}}/g_{\it{N\omega}}$ appearing inside the term in parentheses is a function of $\alpha_V$ or $z$ as it 
fulfills equation (\ref{alphafreecouplingsnew}) or (\ref{zfreecouplings}) respectively. 
We fix $\tilde{g}_{N\omega}$ to the saturation properties of nuclear matter (following Ref. \cite{GlendenningBook}), 
because then it is clear that the EoS of pure nuclear matter is independent of $\alpha_V$ and $z$. 
This is desired, because with or without $\phi$ we wish the properties of nuclear matter to 
be independent of any variations in the hyperon coupling strength relative to the nucleon one. In the TM1 model there is an 
additional $d(\omega^\mu)^4/4$ term in the Lagrangian 
that prevents the mean meson fields from droping out of the equation when substituting as in (\ref{substitutegparagnew}). 
Thus, it becomes necessary to solve the nonlinear equation of motion for the new and old $\omega$ at nuclear saturation density. To fully resolve the 
problem of nonlinear vector meson self-interactions 
we would need to adapt $g_{\it{N\omega}}$ as well as $d$ to the presence of $\phi$ mesons, which requires an additional input
from experiments. 
\section{Results and Discussions}\label{results}
\subsection{Varying the $g_8/g_1$ Ratio $z$}
\begin{figure}
\includegraphics[height=8.6cm,angle=270]{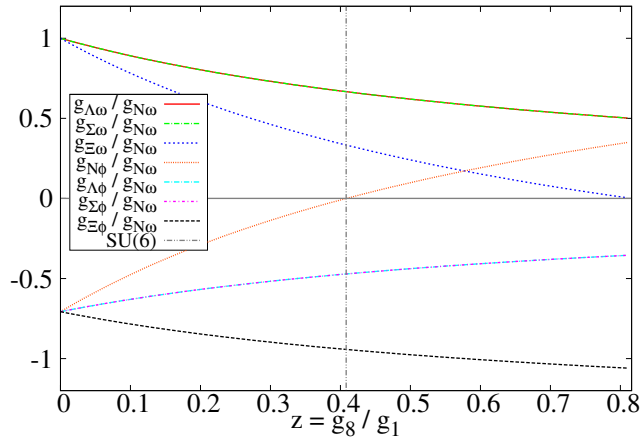} 
\caption{Relative vector meson coupling constants as functions of the $g_8/g_1$ ratio $z$ for fixed $\alpha_V=1$. 
The value $z=1/\sqrt{6}$ corresponds to the SU(6) case.}
\label{figurezfree}
\end{figure}

\begin{figure}
 \includegraphics[width=8.6cm]{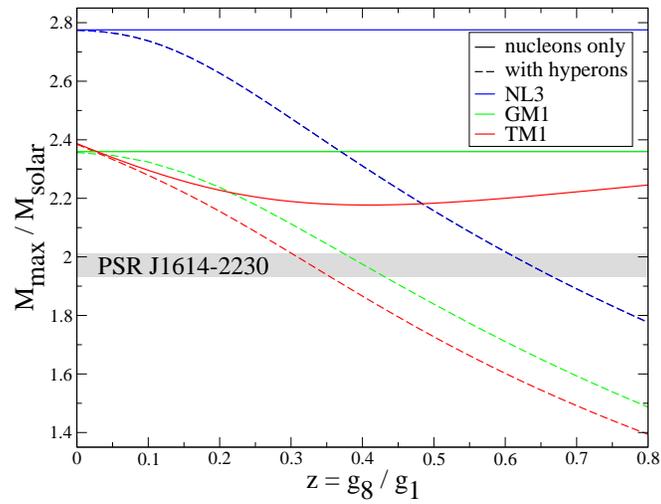}
\caption{Maximum masses as functions of the $g_8/g_1$ ratio $z$ for NL3, GM1 and TM1 parameter sets. 
For each parameterization the case of pure nucleonic matter is also displayed.}
\label{zfreedifferentmodelsfigure}
\end{figure}

We probe the effects 
of the $g_8/g_1$ ratio $z$ on the stiffness of the hadronic EoS. We plot the coupling constants as functions of $z$ in Fig. \ref{figurezfree}, for the 
explicit formulae given in equation (\ref{zfreecouplings}). 
For $z=0$, all coupling constants $g_{\it{B\omega}}$ are the same and similarly are all coupling constants $g_{\it{B\phi}}$ equal. 
This is due to the fact that $z=0$ corresponds to $g_8=0$, 
which results in the equality of the corresponding baryon-meson couplings as the baryons couple only to the flavor
singlet state. 
With increasing $z$, i.e. with increasing contribution from the coupling to the octet $g_8$, 
the resulting couplings all become smaller except for $g_{\it{\Xi\phi}}$. 
At $z=1/\sqrt{6}\approx0.4082$ the SU(6) case is reached where the $\phi$ does not couple to the nucleon. 
Thereafter, for $z>1/\sqrt{6}$ the coupling constant $g_{\it{N\phi}}$ changes its sign so that it is not a
repulsive but now an attractive interaction. Note, that the $\omega$ coupling constants for $\Lambda$ and $\Sigma$ hyperons
are equal for all values of $z$. As anticipated in Sec. \ref{beyondsu6}, we restrict $z$ to the interval $z\in[0:2/\sqrt{6}]$. \\ 
\begin{figure}
 \includegraphics[width=8.6cm]{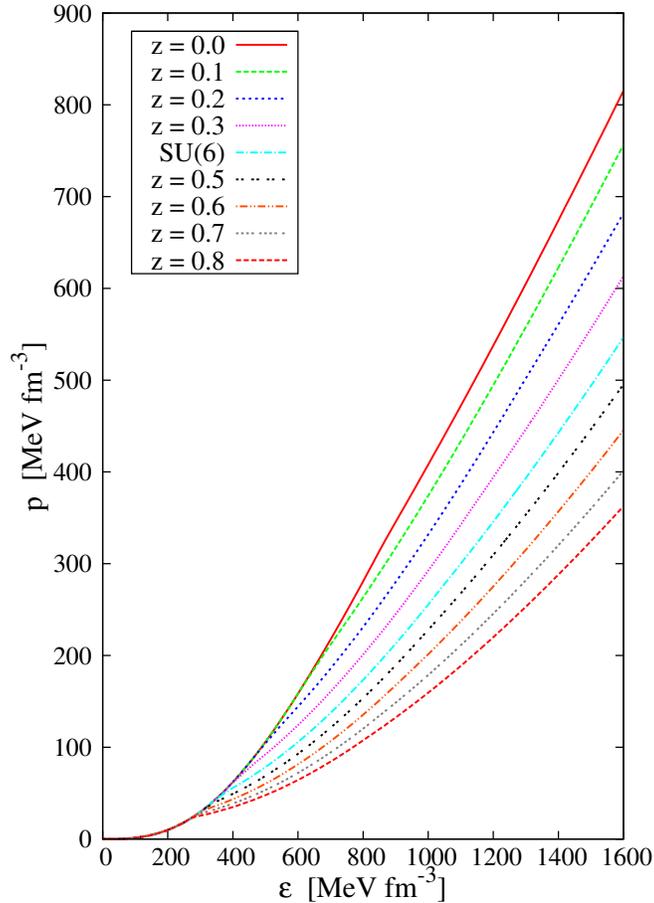}
\caption{EoS for different $g_8/g_1$ ratios $z$ within a nonlinear 
$\sigma-\omega$ model with additional $\phi$ meson and the full baryon octet for GM1 parameter set. The EoS 
get stiffer with decreasing $z$.}
\label{zfreeeosfigure}
\end{figure}
\indent The EoS for $z=0.1,0.2,...,0.8$ are plotted in Fig. \ref{zfreeeosfigure}. 
At first glance it becomes clear that the EoS stiffens with decreasing $z$. 
This can be explained with the help of Fig. \ref{figurezfree} where we had plotted the $z$ dependence of the 
vector meson coupling constants. We noticed that with increasing $z$, all couplings except $g_{\it{\Xi\phi}}$ decrease.
Since the $\Xi$ hyperons only play a subordinate role compared to the neutrons, the increase of $g_{\it{\Xi\phi}}$ does not prevent 
that part of the overall interaction between the baryons, which is mediated by the vector mesons, to become less repulsive. Therefore, the EoS must soften with increasing $z$. 
Together with further decreasing coupling strengths between the vector mesons and the other baryons (except again $g_{\it{\Xi\phi}}$) 
the EoS becomes even softer until $z=2/\sqrt{6}$ is reached. \\
\begin{figure}
\includegraphics[width=08.6cm]{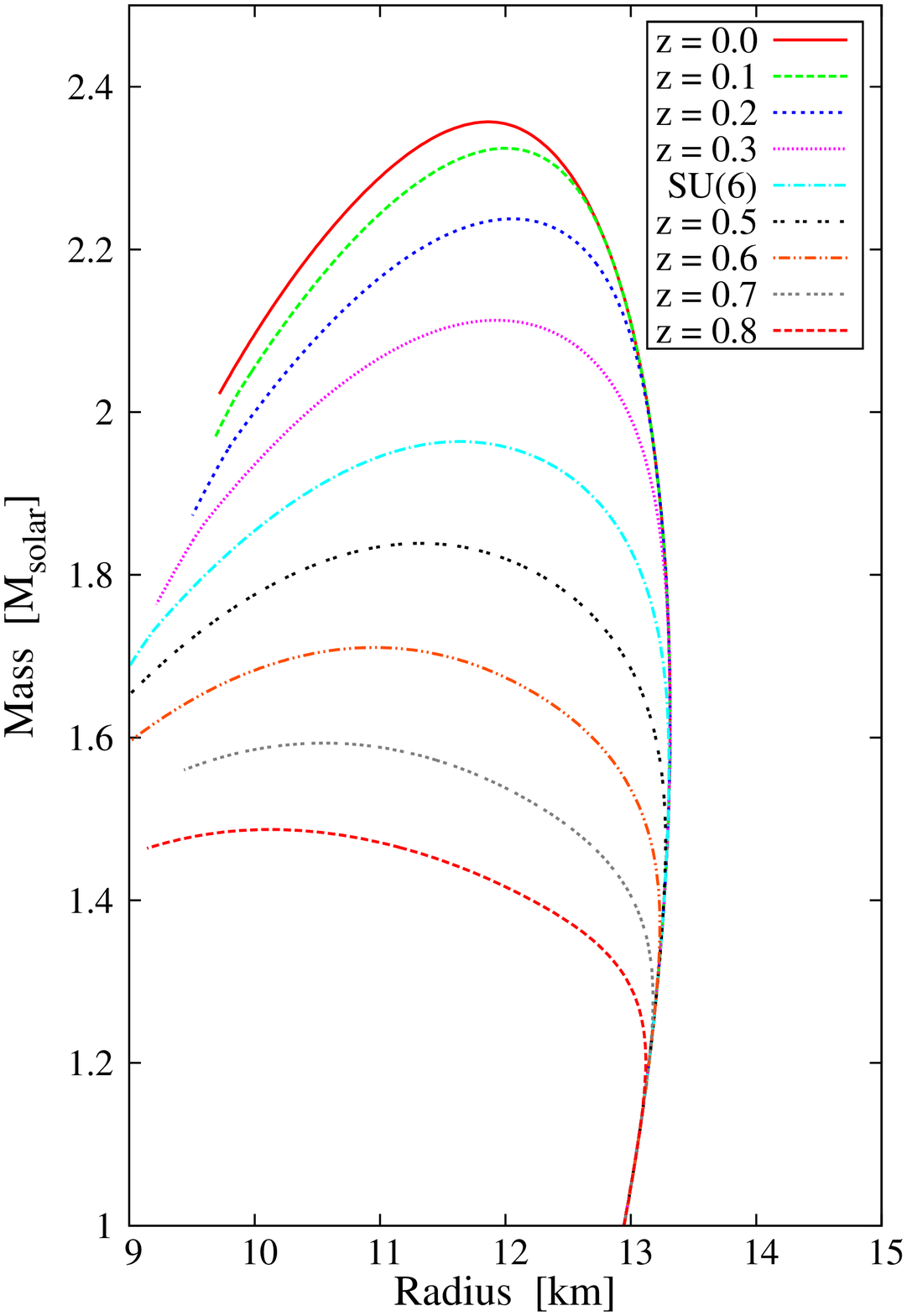}
\caption{Mass-radius relations for the EoS displayed in Fig. \ref{zfreeeosfigure}.
The maximum mass is obtained for the case $z=0$, where all baryons couple to the vector mesons with equal strengths.}
\label{zfreemrfigure}
\end{figure}
\indent In Fig. \ref{zfreemrfigure} we plot the mass-radius relations for the various EoS we just discussed 
in the context of Fig. \ref{zfreeeosfigure}. As expected from the influence of $z$ on the stiffness of the EoS, 
the lowest maximum mass is obtained for $z=2/\sqrt{6}\approx0.8165$, or in the case of Fig. \ref{zfreemrfigure} at $z=0.8$, namely 
$M=1.49M_{\odot}$. The maximum mass grows up to the value $M=2.36 M_{\odot}$ for $z=0$. We notice that the maximum mass of a neutron star in our 
RMF model reacts rather strongly to the variation of $z$: 
over the whole $z$ range the change in the maximum mass is $\Delta_M$=0.87$M_{\odot}$.\\ 
\indent After varying $z$ in rather big steps, we now plot in Fig. \ref{zfreedifferentmodelsfigure} 
the maximum neutron star mass as a continuous function of $z$ for the models GM1, NL3 and TM1 
and for nuclear matter as well as for baryonic matter. 
 As already analyzed in the discussion of Fig. \ref{figurezfree}, 
we see that the branch for nucleonic matter is insensitive to the changes in $z$ for the 
NL3 and GM1 models. For TM1, the quartic self interaction term in the Lagrangian spoils this property since we had to keep the 
corresponding coupling constant $d$ from the SU(6) value of $z$ fixed for the whole $z$ range: 
the maximum masses for pure nucleonic stars therefore depend on the actual $z$ value, 
showing a minimum for the SU(6) case $z\approx0.4082$, namely $M=2.18M_{\odot}$. \\
\begin{figure}
 \includegraphics[width=8.6cm]{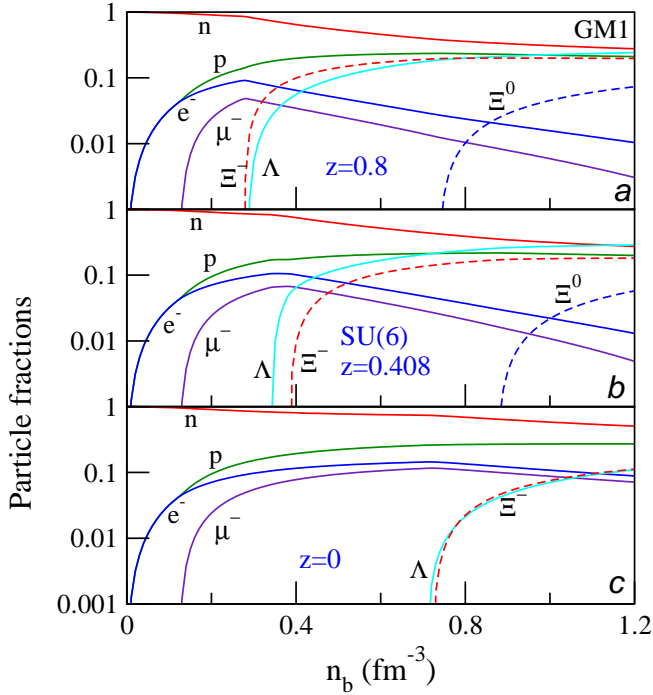}
\caption{Particle fractions for the GM1 parameter set for 3 different values of $z$: $z=0.8$ (a), $z=0.408$ (b) corresponding to the SU(6) case,
and $z=0$ (c). The threshold for the appearance of the hyperons $\Lambda$, $\Xi^-$ and $\Xi^0$ is pushed
to higher densities with decreasing $z$.}
\label{GM1frac}
\end{figure}
Considering the neutron stars containing hyperons, we see in Fig. \ref{zfreedifferentmodelsfigure} that the maximum masses depend on $z$ as 
already observed in Fig. \ref{zfreemrfigure}: 
for the largest $z$ values the maximum masses are the smallest and continually grow with decreasing $z$. 
It is interesting to see that towards $z=0$ the maximum masses of these stars seem to approach the maximum masses of the corresponding pure nucleonic stars 
for all parameter sets studied. 
A look at the particle number fractions for the GM1 parameter set at $z=0.8$ (which we plot in Fig. \ref{GM1frac}(a)) shows that 
the first hyperons to appear in the hadronic matter are the $\Xi^-$ and the $\Lambda$ at total baryon number densities 
of $n_b\approx$ 0.28 ${fm}^{-3}$ and $n_b\approx$ 0.29 ${fm}^{-3}$ respectively, while the $\Xi^0$ appears much later
at $n_b\approx$ 0.76 ${fm}^{-3}$. On increasing $z$ to its SU(6) value (Fig. \ref{GM1frac}(b)), the $\Lambda$ hyperon appears first at 
$n_b\approx$ 0.36 ${fm}^{-3}$, followed by $\Xi^-$ at $n_b\approx$ 0.4 ${fm}^{-3}$ and $\Xi^0$ at
$n_b\approx$ 0.89 ${fm}^{-3}$. At $z=0$ (Fig. \ref{GM1frac}(c)), the threshold of appearance of hyperons is pushed to even higher densities:
 $n_b\approx$ 0.73 ${fm}^{-3}$ for $\Lambda$, $n_b\approx$ 0.74 ${fm}^{-3}$ for  $\Xi^-$ and  $n_b\approx$ 1.38 ${fm}^{-3}$
for $\Xi^0$. Thus, for $z=0$ the neutron stars consist mainly of nuclear matter which is why the maximum masses are so close to those of the pure nucleonic stars. 
For the case of the parameter sets NL3 and TM1, the particle fractions are qualitatively the same as in the GM1 case. In the case of NL3 parameterization, a well-known
instability occurs at high densities when the effective nucleon mass becomes zero \cite{Schaffner96}. The critical density for the 
appearance of the instability depends on the value of the hyperon coupling constants. However for the present investigation, this instability
plays no role as it appears beyond the maximum densities reached in the neutron star interior. 
\subsubsection{Combining $m_N^*$ and $z$ variations}
\begin{figure}
 \includegraphics[width=8.6cm]{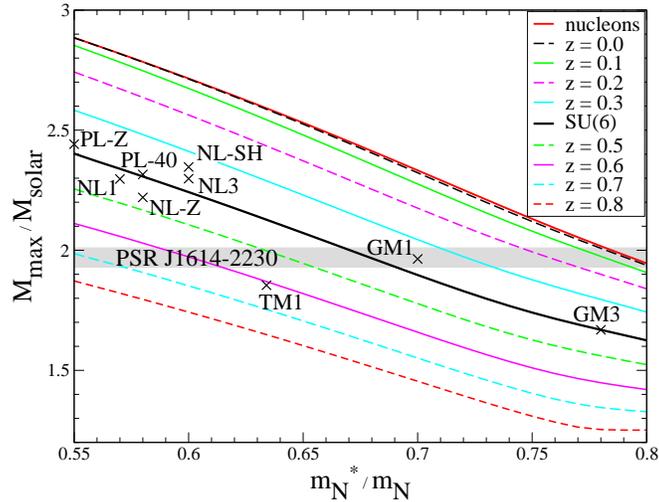}
\caption{Maximum masses of hyperonic neutron stars as functions of effective nucleon mass $m_N^*$ for different values of the $g_8/g_1$ ratio $z$. 
For comparison, a line for nucleonic stars and points to mark RMF sets (e.g. TM1, NL3) corresponding to the SU(6) case are also given.}
\label{figuremeffz}
\end{figure}

The impact of z on the maximum mass of neutron stars is as comparably large as 
the influence of the effective nucleon mass at saturation $m_N^*$ as investigated in our previous study \cite{WsbChtt12}. 
We therefore combine both parameters in a single plot, Fig. \ref{figuremeffz}, where we show the maximum neutron star mass as a 
function of $m_N^*/m_N$ for different $z$ values.
 The incompressibility in this case is fixed to $K=240$ MeV, but the exact value is irrelevant as 
shown in our previous paper \cite{WsbChtt12}. 
We see in Fig. \ref{figuremeffz} that the effective mass has basically the same effect for all $z$ values and $z$ the same effect for all effective masses: 
for fixed $z$, the maximum masses decrease drastically for increasing effective mass. 
For low $z$ values, where the EoS is stiffer than for higher $z$ values, the dependence of the maximum mass on the effective mass is slightly larger: 
the difference along the whole range of $m_N^*/m_N$ is $\approx$ 0.6${M}_\odot$ for $z=0.8$ while for $z=0$ it is $\approx$ 0.9${M}_\odot$. 
The influence of $z$ on the maximum masses is slightly more pronounced than that of the effective masses: 
the difference in the maximum mass between $z=0$ and $z=0.8$ is $\approx$ 0.65${M}_\odot$ for $m_N^*/m_N=0.8$ 
and about 1${M}_{\odot}$ for $m_N^*/m_N=0.55$. 
For comparison, we also plot the maximum masses of purely nucleonic neutron stars, and also mark in the figure the points
corresponding to the SU(6) case, for several other RMF sets fitted to properties of nuclei e.g., TM1, NL3 or NL-SH \cite{NL3,Sugahara94,NL-Z,PL-Z,NL-SH}. 
In Fig. \ref{zfreedifferentmodelsfigure} the masses of baryonic stars were found to 
approach the limit of purely nucleonic stars for decreasing values of z. 
With Fig. \ref{figuremeffz} we can now visualize that the maximum masses in these two cases differ slightly ($\Delta_M<0.01M_\odot$).\\
\indent We close this section by concluding that a maximum mass of at least 1.97 $\pm$ 0.04${M}_{\odot}$ requires very small values of $z$ 
for large effective masses ($z\leq0.1$ at $m_N^*/m_N=0.8$), $z$ values around SU(6) for effective masses close to $m_N^*/m_N \approx 0.7$ and 
very low effective masses for large $z$ values ($m_N^*/m_N < 0.58$ for $z=0.7$). 
We note, that $z$ values close to the maximum of $z=2/\sqrt{6}$ are now allowed configurations within the plot range: 
the maximally allowed $z$ value for the investigated model is $z < 0.77$ at $m_N^*/m_N=0.55$. 
\subsection{Varying the F/(F+D) ratio $\alpha_V$}
\begin{figure}
  \includegraphics[height=8.6cm,angle=270]{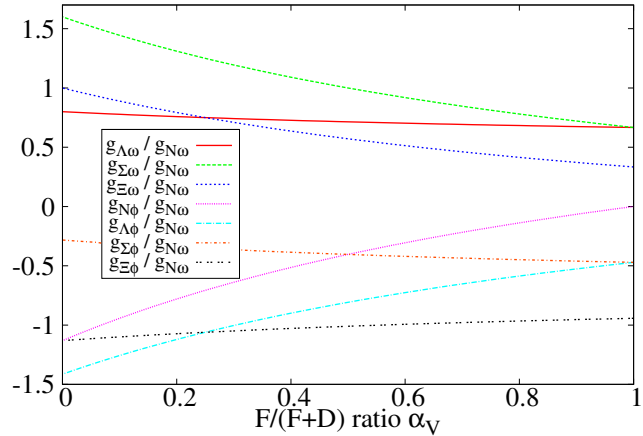}
\caption{Vector meson coupling constants as functions of the $F/(F+D)$ ratio $\alpha_V$. The $g_8/g_1$ ratio is fixed to 
its SU(6) value $z=1/\sqrt{6}$ and ideal mixing is assumed. 
The SU(6) case is given by $\alpha_V=1$.}
 \label{figurealphafreeB}
 \end{figure}

\begin{figure}
\includegraphics[width=8.6cm]{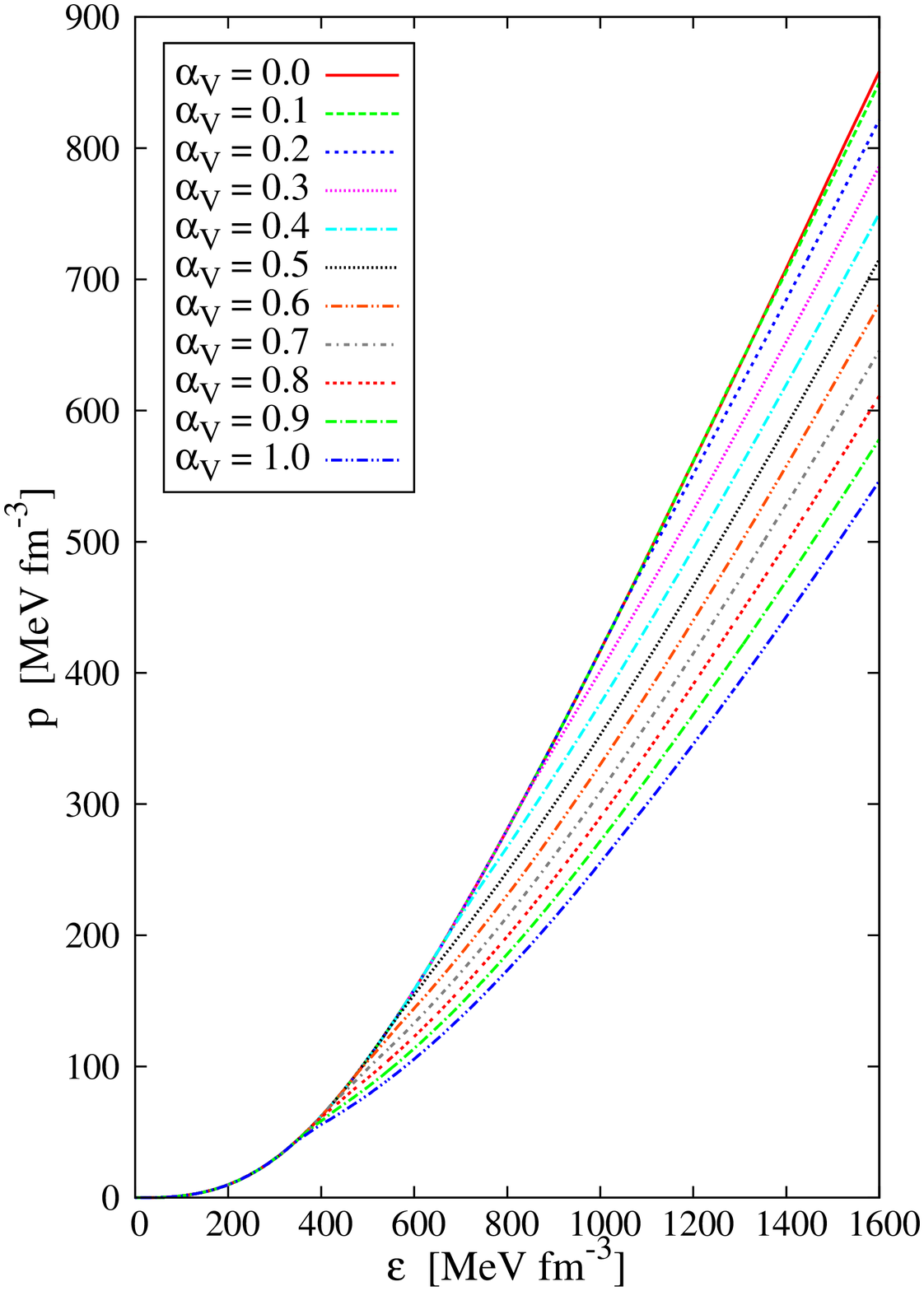} 
\caption{EoS for ``model $\sigma \omega \rho \phi$`` in GM1 parameterization for different values of $\alpha_V$. $z$ is fixed to its SU(6) value $z=1/\sqrt{6}$ and ideal mixing is assumed.
The EoS become stiffer with decreasing $\alpha_V$.}
\label{figurealphaeosB}
\end{figure}

\begin{figure}
\includegraphics[width=8.6cm]{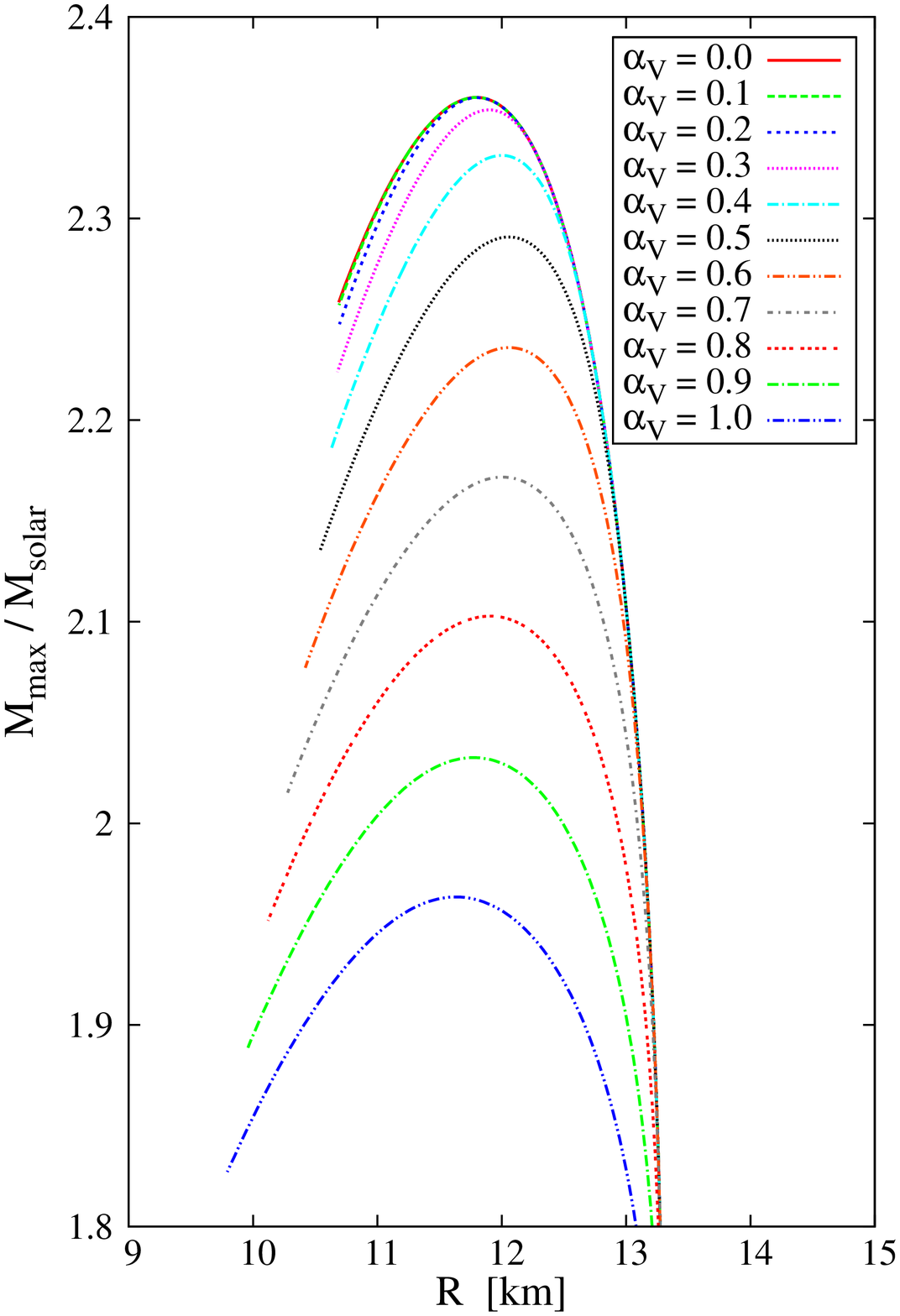} 
\caption{Mass-radius relations obtained from the EoS in Fig. \ref{figurealphaeosB}.}
\label{figurealphamrB}
\end{figure}

After systematically investigating a wider $z$ range we repeat the formalism for the $\alpha_V$-ratio 
in the present section. 
The $F/(F+D)$ ratio is by definition restricted to the interval $\alpha_V\in[0;1]$, where the lower bound corresponds to a pure D-type 
coupling and the upper limit (i.e. the SU(6) value) corresponds to a pure F-type coupling.  
Analogous to the case studied above, we adopt ideal mixing as well as 
a $g_8/g_1$ ratio fixed to its SU(6) value $z=1/\sqrt{6}$ and we allow for the $\phi$ meson to couple to the nucleon. 
We plot the coupling strengths in Fig. \ref{figurealphafreeB} where we vary $\alpha_V$ between 0 and 1. 
Note that the coupling constants $g_{\omega \Lambda}$ for $\Lambda$ hyperons do not change considerably and that all vector couplings remain
repulsive (do not change their sign).\\
\indent The ratio of the F- and D-type couplings can be continuously varied between the two extremes of a pure F- and a pure D-type coupling. 
We see that from $\alpha_V=1$ down to $\alpha_V=0$ all couplings become stronger except for $g_{\it{\Sigma\phi}}/g_{\it{N\omega}}$. 
Since the $\Sigma$ hyperons have but very little influence on the EoS up to neutron star densities, 
we can expect that the EoS become stiffer with decreasing $\alpha_V$. This is exactly what we find 
in Fig. \ref{figurealphaeosB} where we plot the EoS for several values of 
$\alpha_V$ using "model $\sigma \omega \rho \phi$" and GM1 parameter set.\\
\indent The stiffness depends monotonously on $\alpha_V$ and we get the softest EoS for the SU(6) case $\alpha_V=1$, 
i.e. a pure F-type coupling, while the stiffest EoS is obtained for $\alpha_V=0$ which corresponds to the pure D-type 
coupling of the baryon and meson multiplets. 
The EoS for $\alpha_V\leq0.2$ appear to be indistinguishable at neutron star densities. 
This is evident in Fig. \ref{figurealphamrB}, where we plot the mass-radius relations corresponding to the EoS from Fig. \ref{figurealphaeosB}: 
for the values $\alpha_V=0.0-0.2$ the maximum masses but also the radii of the corresponding stars coincide (M$_{\it{max}}=2.36$M$_\odot$, R$=11.8$ km). 
We note, that this value of the maximum mass is also obtained for the purely nucleonic case (compare e.g. Fig. \ref{zfreedifferentmodelsfigure}). 
Thus, for the "model $\sigma \omega \rho \phi$" the nuclear matter limit is reached below $\alpha_V < 0.2$ in the case of GM1, GM3 and NL3 parameter sets, 
while for the very stiff PL-Z EoS (having effective mass $m^*_N/m_N \simeq 0.55$) \cite{PL-Z} pure nucleonic stars are already obtained for $\alpha_V < 0.3$. \\
\indent In this way, for hyperonic stars the limit of nucleonic stars is continuously approached 
for decreasing values of the $g_8/g_1$ 
ratio $z$, or for decreasing values of the $F/(F+D)$ ratio $\alpha_V$ away from the SU(6) value, respectively. 
\\ \\
\begin{figure}
 \includegraphics[width=8.6cm]{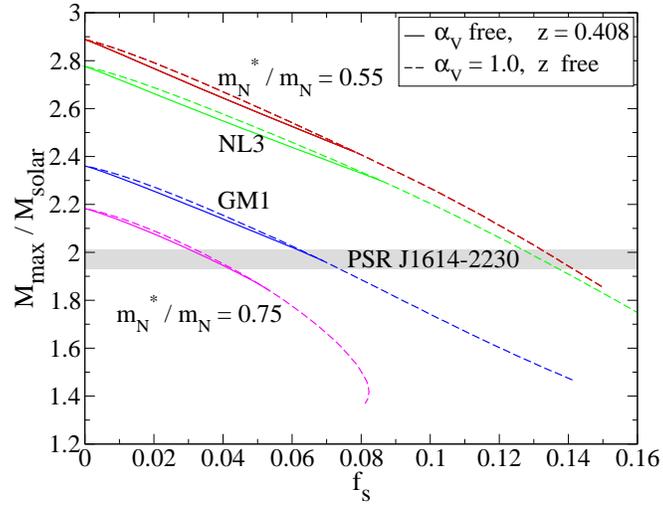}
\caption{Maximum masses of neutron stars as functions of strangeness fraction $f_s$ in the neutron star core for four different EoS. 
The maximum mass decreases linearly with the strangeness fraction approximately as $M_{max}(f_s) = M_{max}(0) - 0.6 M_{\odot} (f_s/0.1)$.}
\label{mmax_fs}
\end{figure}
\indent To generalize our findings, we plot in Fig. \ref{mmax_fs} the maximum masses of neutron stars as a function of strangeness fraction
$f_s$ (the number of strange quarks divided by the total number of quarks) for four different EoS: for $m_N^*/m_N$ = 0.55, 0.75 and for the NL3 ($m^*_N/m_N = 0.6$) and 
GM1 ($m^*_N/m_N = 0.7$) parameter sets, by varying $\alpha_V$
(solid lines) and $z$ (dashed lines). It is evident from the figure that on decreasing $\alpha_V$ or $z$, the strangeness
fraction in the core decreases, and there is a corresponding increase in the maximum mass of the star. At zero strangeness
fraction the nucleonic limit is reached, and this corresponds to the highest value of the maximum mass. For $m^*_N/m_N \lessapprox 0.7$, the relation between
the maximum mass of the star and its strangeness fraction can be fitted linearly according to the formula:
\begin{equation}
\frac{M_{max}}{M_{\odot}} = \frac{M_{max}  (f_s = 0)}{M_{\odot}} - c \left(\frac{f_s}{0.1}\right)~,
\end{equation}
where $c\approx0.6M_{\odot}$. This has interesting consequences when we use these results to predict the maximally allowed
strangeness fraction in maximum mass neutron stars. In an associated study \cite{TolosSagert}, we applied the results
from measurements of sub-threshold kaon production in heavy-ion collisions (HIC) to study the implications on
neutron star properties. It was found that the heavy-ion data and causality imply a firm upper limit on
maximum mass of compact stars of 3 solar masses. Substituting this value into the derived formula above,
we can show that the strangeness fraction in a 2 $M_{\odot}$ star cannot be more than:
\begin{eqnarray}
f_s^{max} &=& \frac{M_{max}^{HIC} - M_{max}^{obs}}{6 M_{\odot}} \nonumber\\
&=& \frac{3 M_{\odot} - 2 M_{\odot}}{6 M_{\odot}} = 0.17 ~.
\end{eqnarray}
\indent We also point out that for the EoS compatible with the observed mass limit, the hyperon fraction in
a canonical $1.4M_{\odot}$ star is zero. Only for small effective nucleon masses and values of $z$ above the SU(6) value can 
a hyperon fraction of less than $0.5\%$ be reached.\\
\indent We have now found that we can reach the nuclear matter limit starting from a RMF model including hyperons and continuously changing 
the model parameters, instead of making a discrete ``on/off'' decision about whether to include or exclude hyperons. 
Instead, we should take the position of saying that one can only exclude a RMF parameter set 
as soon as the corresponding maximum mass for nucleonic stars is incompatible with observations. 
As long as the observational mass limit is below the nucleonic mass limit of the model, it is possible to have hyperons
in the core of maximum mass neutron stars. 
\section{Summary}\label{summary}
\indent We investigated the conditions for the existence of hyperons in massive neutron stars. We went beyond the spin-flavor SU(6) quark 
model for determining the vector meson couplings to the more general relations from flavor SU(3), 
where we varied the $g_8/g_1$ ratio $z$ and the F/(F+D) ratio $\alpha_V$ while assuming ideal mixing. \\
\indent We fixed either $z$ or $\alpha_V$ to their SU(6) values and allowed for a non-vanishing $N-\phi$ coupling. 
We found within "model $\sigma \omega \rho \phi$", that decreasing $z$ below its SU(6) value leads to stiffer EoS and larger maximum masses. The most massive stars 
were obtained for $z=0$, where only the meson singlet couples to the baryon octet, meaning that all 
vector couplings take the same value. 
We have seen that at $z=0$ the masses of the stars come very close to the case of purely nucleonic neutron stars.  \\
\indent Keeping $z$ fixed at its SU(6) value and varying $\alpha_V$ has even a greater impact on the maximum masses of neutron stars: 
the lowest value is given for the SU(6) case $\alpha_V=1$. Towards $\alpha_V=0$ the maximum masses increase monotonically. 
In both cases, i.e. on lowering $z$ and on lowering $\alpha_V$, we found that the combined repulsive interactions mediated by the $\omega$ and $\phi$ mesons 
cause the hyperons to appear at successively higher densities so that hyperons appear only in a small core 
($z=0$, $\alpha_V=1$) of the maximum mass star or not at all ($z=1/\sqrt{6}$, $\alpha_V=0$). 
Thus, the EoS becomes stiffer and the strangeness fraction of neutron stars decreases 
with decreasing values of $\alpha_V$ or z until the limit is reached where 
the whole sequence of neutron stars contains only pure nucleonic stars. 
There is a smooth transition between strange hadronic and nucleonic stars as well as between their corresponding 
maximum masses where formerly there was only an explicit inclusion/exclusion of hyperons in the construction of the model. 
This finding holds for all investigated parameter sets, but it should be stressed that the inclusion of the $\phi$ is vital for reaching the nuclear limit. 
We find that the maximum mass decreases linearly with the strangeness fraction $f_s$ of the neutron star core as $M_{max}(f_s) = M_{max}(0) - 0.6 M_{\odot} (f_s/0.1)$,
independent of the chosen nuclear EoS and the adopted vector baryon-meson coupling scheme. 
For instance, given the mass of $3M_\odot$ for a purely nucleonic neutron star, 
the strangeness fraction for a neutron star with an observed mass of $1.97M_\odot$ would then be below $f_s \lessapprox  17 \%$.\\ \\
\section*{Acknowledgments}
D.C. and S.W. are thankful to Pawel Haensel, J\'er\^ome Margueron, Micaela Oertel and Wynn Ho for
fruitful discussions and suggestions. 
J.S.-B. is supported by the DFG through 
the Heidelberg Graduate School of Fundamental Physics. 
D.C. acknowledges the support from the Alexander von Humboldt foundation. 
S.W. is supported by the state of Baden-W\"urttemberg through a LGFG stipend. 
This work is supported by BMBF under Grant No. 
FKZ 06HD9127, by the Helmholtz Alliance HA216/EMMI, and by CompStar, 
a research networking program of the European Science Foundation.\\ \\
\bibliographystyle{utphys}
\providecommand{\href}[2]{#2}\begingroup\raggedright\endgroup
\end{document}